\documentclass[twocolumn, superscriptaddress, amsmath, amssymb, aps, prl, showpacs, floatfix]{revtex4-2}

\usepackage{graphicx}
\usepackage{dcolumn}
\usepackage{bm}
\usepackage[colorlinks,linkcolor=blue,citecolor=blue,anchorcolor=blue]{hyperref}
\usepackage[mathlines]{lineno}
\usepackage{setspace}
\usepackage{xcolor}
\usepackage{amssymb}
\usepackage{textcomp} 
\usepackage{gensymb} 
\usepackage{soul} 
\usepackage{etoolbox} 

\newcommand{\Yb}{{}$^{171}$Yb}

\newcommand{\ket}[1]{\left|  #1 \right\rangle}

\newcommand{\aver}[1]{\ensuremath{\langle {#1} \rangle}}

\definecolor{plotgreen}{RGB}{0,150,0}

\begin{document}

\title{Increased Atom-Cavity Coupling through Cooling-Induced Atomic Reorganization}

\author{Chi Shu}
\thanks{These authors contributed equally}
\affiliation{Department of Physics, MIT-Harvard Center for Ultracold Atoms and Research Laboratory of Electronics, Massachusetts Institute of Technology, Cambridge, Massachusetts 02139, USA}
\affiliation{Department of Physics, Harvard University, Cambridge, Massachusetts 02138, USA}
\author{Simone Colombo}
\altaffiliation{Current address: Department of Physics, University of Connecticut, Storrs, Connecticut 06269, USA}
\affiliation{Department of Physics, MIT-Harvard Center for Ultracold Atoms and Research Laboratory of Electronics, Massachusetts Institute of Technology, Cambridge, Massachusetts 02139, USA}
\author{\!\!$^{,*}$ Zeyang Li}
\altaffiliation{Current address: Department of Applied Physics, Stanford University, Stanford, California 94305, USA}
\affiliation{Department of Physics, MIT-Harvard Center for Ultracold Atoms and Research Laboratory of Electronics, Massachusetts Institute of Technology, Cambridge, Massachusetts 02139, USA}
\author{\!\!$^{,*}$ Albert Adiyatullin}
\altaffiliation{Current address: Quandela, 7 Rue Leonard de Vinci, 91300 Massy, France}
\affiliation{Department of Physics, MIT-Harvard Center for Ultracold Atoms and Research Laboratory of Electronics, Massachusetts Institute of Technology, Cambridge, Massachusetts 02139, USA}

\author{Enrique Mendez}
\affiliation{Department of Physics, MIT-Harvard Center for Ultracold Atoms and Research Laboratory of Electronics, Massachusetts Institute of Technology, Cambridge, Massachusetts 02139, USA}

\author{Edwin Pedrozo-Pe\~{n}afiel}
\affiliation{Department of Physics, MIT-Harvard Center for Ultracold Atoms and Research Laboratory of Electronics, Massachusetts Institute of Technology, Cambridge, Massachusetts 02139, USA}

\author{Vladan Vuleti\'{c} }
\email{vuletic@mit.edu}
\affiliation{Department of Physics, MIT-Harvard Center for Ultracold Atoms and Research Laboratory of Electronics, Massachusetts Institute of Technology, Cambridge, Massachusetts 02139, USA}

\date{\today}
\begin{abstract}
The strong coupling of atoms to optical cavities can improve optical lattice clocks as the cavity enables metrologically useful collective atomic entanglement and high-fidelity measurement. To this end, it is necessary to cool the ensemble to suppress motional broadening, and advantageous to maximize and homogenize the atom-cavity coupling. 
We demonstrate resolved Raman sideband cooling via the cavity as a method that can simultaneously achieve both goals. 
In $200$~ms, we cool \Yb~atoms to an average vibration number $\aver{n_x}= 0.23(7)$ in the tightly binding direction, resulting in $93\%$ optical $\pi$-pulse fidelity on the clock transition ${}^1S_0 \to {}^3P_0$. During cooling, the atoms self-organize into locations with maximal atom-cavity-coupling, which will improve quantum metrology applications.
\end{abstract}
                            
\maketitle

\begin{figure}
    \centering
    \includegraphics[width =\columnwidth]{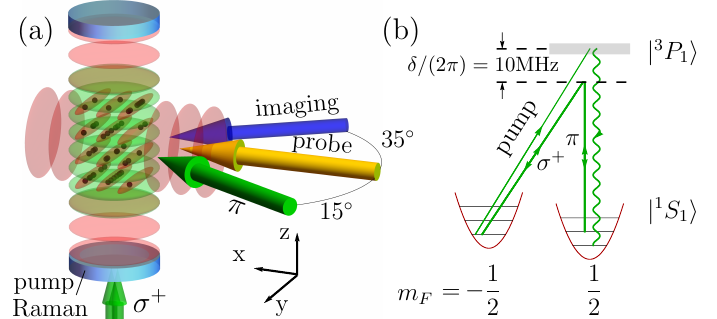}
    \caption{Raman sideband cooling in an optical cavity. (a) A cavity along the vertical $z$ direction supports both a trapping optical lattice at 759~nm and light near the $\ket{^1S_0,m_F=\frac{1}{2}} \rightarrow \ket{{^3P}_1,m_F=\frac{3}{2}}$ transition at 556 nm for optical pumping, and serving as one leg of the Raman transition. An additional $\pi$-polarized Raman beam is applied in the $xy$ plane at an angle of 15$\degree$ relative to the $x$ axis. The laser beam driving the $\ket{^1S_0} \rightarrow \ket{{^3P}_0}$ optical-clock transition propagates along $x$. Time-of-flight imaging on the $^1S_0 \rightarrow {^1P}_1$ transition at 399~nm is performed at 35$\degree$ against the $x$ direction in $xy$ plane. (b) With $B_z$ = 13.5 G, Zeeman splittings in ${^3P}_1$ and $^1S_0$ are 20~MHz and 10~kHz respectively. Raman beams are red-detuned by 10 MHz to the $\ket{^1S_0,m_F=\frac{1}{2}} \rightarrow \ket{{^3P}_1,m_F=\frac{1}{2}}$ transitions with resonant two-photon condition between the two ground states.} 
    \label{fig1}
\end{figure}

Ultracold atomic ensembles in optical cavities constitute a versatile platform for a wide range of applications, from generating nonclassical states of light~\cite{Sayrin2011,Vlastakis2013,Hacker2016,Ramette2022}, to mediating atom-atom interaction for quantum metrology \cite{Leroux2010,Schleier-Smith2010a,Bohnet2014a,Cox2016a,Hosten2016,Hosten2016a,Braverman2019,Zhao2021}, quantum information science~\cite{Pellizzari1995,Duan2004,Kimble2008,Weedbrook2012}, and quantum many-body physics~\cite{Ritsch2013,Hung2016,Leonard2017a,Leonard2017b,Vaidya2018,Guo2019a,Guo2019b,Morales2019,Davis2019,Bentsen2019a,Bentsen2019b,Park2022,Langenfeld2021}.
Two technical challenges in such systems are the inhomogeneous coupling of the atoms to the light mode in standing-wave cavities, and achieving sufficiently low temperatures to suppress thermal noise and motional dephasing.

The coupling can be made homogeneous by removing of weakly coupled atoms~\cite{wu2021sitedependent} at the expense of reduced atom number. Alternatively, one can use wavelength-commensurate trapping and interaction optical lattices~\cite{Lee2014a}. However, the latter is not possible in optical lattice clocks that require a particular (magical) trapping wavelength~\cite{Katori2003}. For the cooling, direct laser cooling to Bose-Einstein condensation on a narrow transition \cite{Stellmer2013,Chen2022} and by Raman cooling in alkali metal atoms \cite{Hu2017,Solano2019,Vendeiro2022} have also demonstrated the ability to cool to quantum degeneracy with rather simple experimental setups. In these systems, thermalization through elastic collisions occurs during the cooling. It is unclear whether these methods can be applied to atomic species with a very small elastic collision cross section, such as \Yb.

In this Letter, we report a cavity-light-assisted two-photon Raman sideband cooling method for \Yb~atoms that simultaneously cools to the quantum ground state in the tightly confined directions in a magical-wavelength trap, and reorganizes the atoms along the cavity axis to achieve a stronger and uniform coupling to the cavity. The attained low mean vibrational quantum number $\aver{n_x}=0.23(7)$ enables high-fidelity Rabi oscillations on the ${}^1S_0 \rightarrow {}{^3P}_0$ optical-clock transition, which in combination with the cavity enables entanglement-induced metrological gain \cite{pedrozo2020entanglement}. Due to the very small elastic collision cross section of \Yb, the temperatures along the tightly confined direction ($T_x=1.8(2) \mu$K) and the weakly confined direction ($T_y=8(3) \mu$K) remain decoupled even at atomic densities of $n \sim 10^{11}$~cm$^{-3}$. One feature of the cooling is that the atoms also reorganize along the cavity towards the trapping positions with larger coupling to the cavity, thereby increasing the effective single-atom cooperativity $\eta$ to within 5\% of its maximum value. At the same time, the phase space density (maximum occupation per quantum state) increases to $\textrm{PSD}=0.013(3)$, bringing the system fairly close to quantum degeneracy. 

Atoms are first loaded into a bi-color mirror magneto-optical trap (MOT)~\cite{Kawasaki2015} located inside the optical cavity. By changing the bias magnetic fields, we adjust the MOT to overlap well with the vertical ($z$) cavity mode and a one-dimensional standing-wave optical lattice in the $x$-direction with a waist of 27~$\mu$m. The lattice operates at the magic wavelength $\lambda=759$~nm for the $\ket{^1S_0}\rightarrow \ket{^3P_0}$ clock transition, and has a trap depth $U_x/h = 460$~kHz.

We then turn off the MOT beams and send a second magic-wavelength trapping beam into the cavity to generate an optical lattice along the $z$-direction with a waist of 130~$\mu$m and trap depth $U_z/h=2$ MHz at the atoms' position. The corresponding vibration frequencies of atoms in the two-dimensional (2D) optical lattice are $\{\omega_x/(2\pi), \omega_y/(2\pi), \omega_z/(2\pi)\}$=$\{ 60,1,130\}$~kHz. 
To remove any atoms outside the 2D trap overlap region, the optical lattice along $x$ is adiabatically ramped down, held off for 50~ms, and then slowly ramped back up in 15~ms, followed by a ramping down and back up of the cavity trap light in 32~ms. 
This procedure results in a cloud of $\sim 500$ $^{171}$Yb atoms with root-mean-square (rms) sizes of $16 \mu$m and $4.8 \mu$m along the $z$ and $x,y$ directions, respectively. At this point, the peak local density and peak phase space density in the two-dimensional lattice are $n_0=3\times 10^{11}$~cm$^{-3}$ and $\textrm{PSD}=2\times10^{-4}$, respectively.

Both the Raman coupling and the optical pumping necessary for the Raman sideband cooling \cite{Hamann1998,Vuletic1998,Hu2017} are accomplished with a laser near the $^{1}S_{0}\rightarrow{^{3}P}_{1}$ transition. 
The optical pumping is performed with a $\sigma^+$-polarized beam along the cavity that is resonant with the $\ket{^1S_0,m_F=-\frac{1}{2}} \rightarrow \ket{{^3P}_1,m_F=+\frac{1}{2}}$ transition, while the Raman coupling uses two beams detuned from the $\ket{^1S_0,-\frac{1}{2}} \rightarrow \ket{{^3P}_1,+\frac{1}{2}}$ transition by $\Delta/(2\pi) = -10$~MHz, one $\sigma^+$-polarized beam along the cavity, and a $\pi$-polarized beam in the $xy$ plane (see Fig.~\ref{fig1}). With a B field of $13.5$~G along the $z$ axis, the relative detuning of those two beams is chosen to match the $\ket{^1S_0, m_F=\frac{1}{2}, n_x} \rightarrow\ket{^1S_0, m_F=-\frac{1}{2}, n_x-1}$ transition between the two ground states that reduces the vibrational quantum number $n_x$ by one, and hence the motional energy by $E/h= 60$~kHz. The optical pumping back to the $\ket{m_F=-\frac{1}{2}}$ state heats the atom on average by $2E_{rec}/h = 7.4$~kHz, where $E_{rec}$ is the recoil energy for the $^1S_0 \rightarrow ^3P_1$ transition. 
After cooling for a variable time (1-1000~ms), we extinguish the Raman beams 5~ms before the optical pumping light in order to initialize the atoms in the $\ket{^1S_0, m_F=\frac{1}{2}}$ level. 
After the cooling process, the optical lattice along the $z$ direction is ramped down to 30\% of its initial power to reduce the photon scattering by the trap light. 

\begin{figure}
     \includegraphics[width =\columnwidth]{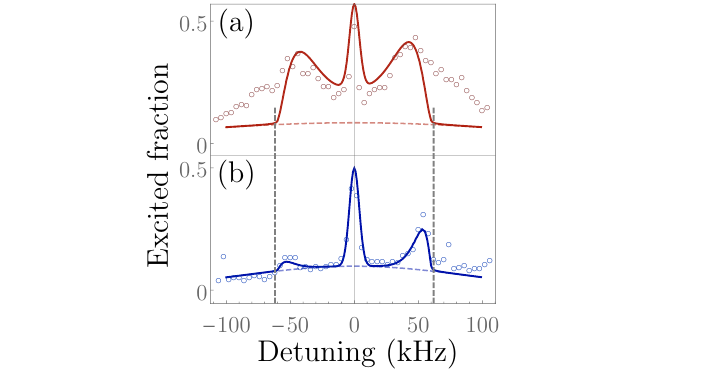}
    \caption{Clock state excitation spectroscopy ($\ket{^1S_0} \rightarrow \ket{{^3P}_0}$) in the 2D optical lattice (a) before and (b) after 200~ms of cooling. Clock pulse lengths of (a) 5~ms and (b) 20~ms are applied. At high temperature, there is a large Doppler broadened background of atoms that are only trapped by the intracavity light. At low temperatures, the vibrational sidebands in the $x$ lattice are prominent, and the red sideband is suppressed. Assuming a thermal distribution, the fitted temperatures of the cloud are $T_{x,i} = 20(2) \mu$K before cooling and $T_{x,f} = 1.8(5) \mu$K after cooling, with mean vibrational occupation numbers  $\aver{n_{x,i}} = 6.1(7)$ and $\aver{n_{x,f}} =0.23(4)$ respectively. }
    \label{Fig:Clock}
\end{figure}

\begin{figure}
    \centering
    \includegraphics[width =0.97\columnwidth]{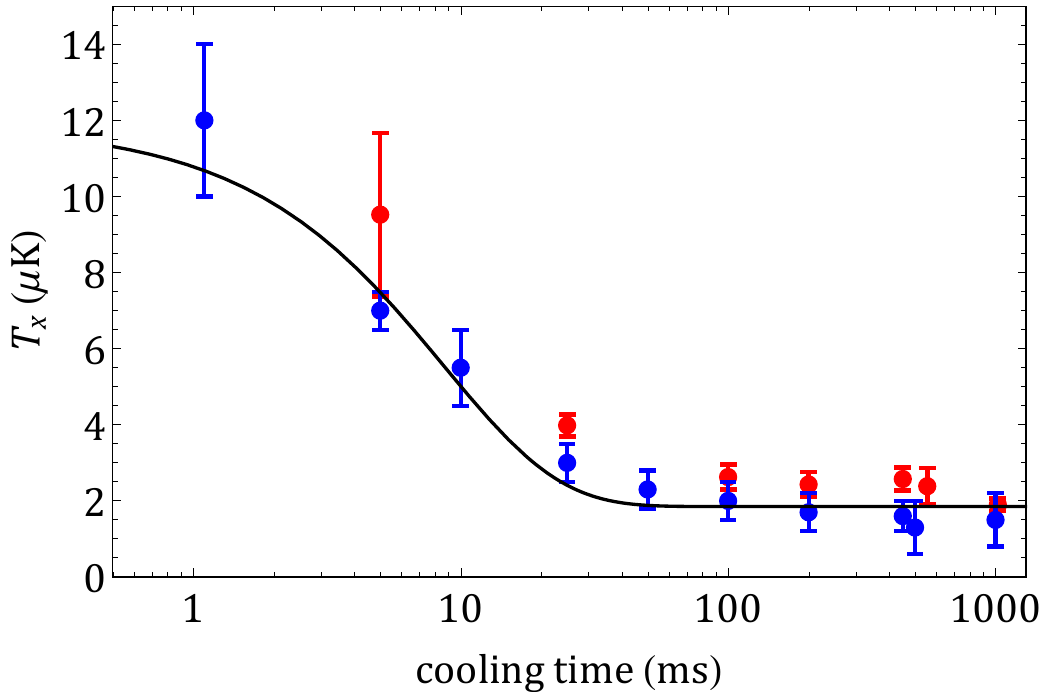}
    \caption{Temperature as a function of cooling time. Blue data represent the temperature obtained by sideband analysis (see Fig:~\ref{fig:Temperature}) while red data are obtain by fitting the Rabi flopping to the theory from~\cite{Blatt2009PRA}. The solid line is an exponential decay fit of the blue data, with a cooling time constant of $9(1)\,$ms and a final temperature of $1.8(5)\mu$K.}
    \label{fig:Temperature}
\end{figure}

The atomic temperature is determined by spectroscopy on the clock transition $^{1}S_{0}\to{^{3}P}_{0}$ using a $\pi$-polarized laser beam traveling along the $x$ direction. This laser is stabilized to an ultralow-expansion cavity and measures the population of vibrational states via sideband spectroscopy~\cite{Blatt2009PRA}, as shown in  Fig.~\ref{Fig:Clock}. The broad Gaussian background underlying the discrete sideband spectrum is attributed to the Doppler profile of floating atoms that are not confined in a single site of the lattice along $x$, but are confined in the intracavity trap with a much larger trap depth. 
Before the cooling, the blue and red sidebands have similar size, which indicates a mean vibrational quantum number $n_x \gg 1$, while the Doppler background in the spectrum implies a large portion of atoms that are not confined in the $x$ lattice. 
After 200~ms of Raman sideband cooling, a large sideband imbalance is observed, implying that the atoms are cooled to near the vibrational ground state with $\aver{n_x}=0.23(7)$. 
The reduced residual Doppler background indicates that atoms originally not confined by the $x$ lattice are cooled to the vibrational ground state as well. 
Fig.~\ref{fig:Temperature} shows a fast initial cooling within the first 10~ms, followed by a slower temperature decrease as the atoms are cooled into the ground state.

\begin{figure}
    \centering
    \includegraphics[width =\columnwidth]{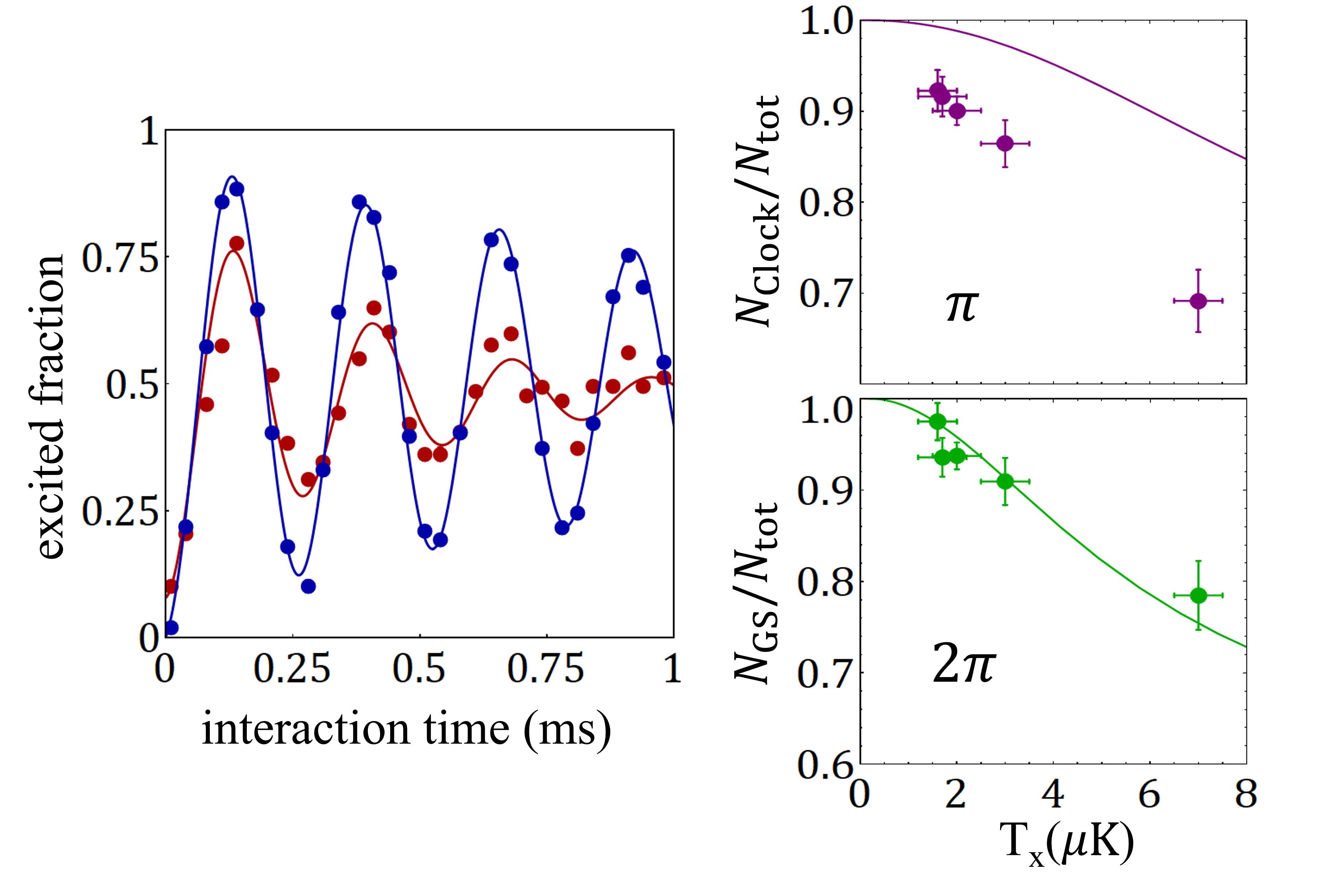}
    \caption{Rabi flopping on the clock transition $\ket{^1S_0} \rightarrow \ket{^3P_0}$ for different temperatures ($T=1.8~\mathrm{\mu K}$, blue datapoints, and $T=10~\mathrm{\mu K}$, red datapoints). The deviation from the theory \cite{Blatt2009PRA} for the $\pi$ pulse (purple points, b) is due to the fraction of atoms that are not confined in the 2D lattice. This fraction increases when the temperature is higher. The $2\pi$ pulse (c) is insensitive to those atoms.}
    \label{fig:Rabi}
\end{figure}

After the cooling, the optical Rabi oscillation experiences a much smaller dephasing than before the cooling, resulting in strongly improved coherent transfer to the excited clock state $\ket{^3P_0}$ (Fig.~\ref{fig:Rabi}a).
After 200~ms of cooling, the transfer fidelity reaches 0.93(3). 
The remaining infidelity can be explained by the residual population of vibrational excited states, which experience different Rabi frequencies $\Omega_{m}$ on the $\ket{^1S_0,n_z=m} \to \ket{{}^3P_0,n_z=m}$ vibrational transition, given by $\Omega_{m} = \Omega_0 e^{-\eta_x^2/2}L_n(\eta_x^2)$. Here $\Omega_0$ is the Rabi frequency for the vibrational ground state, $\eta_x = 0.24(1)$ is the Lamb-Dicke parameter for our lattice depth, and $L_n$ is the Laguerre polynomial. 
When we compare the $\pi$-pulse and $2\pi$-pulse fidelities (Fig. \ref{fig:Rabi}a,b) to each other and to a model~\cite{Blatt2009PRA}, we see that the $\pi$-pulse fidelity is lower and deviates more from the model due to the atoms that are not confined by the $x$ lattice, whereas the $2\pi$-pulse is insensitive to those atoms. 

\begin{figure*}[!ht]
\includegraphics[width=\textwidth]{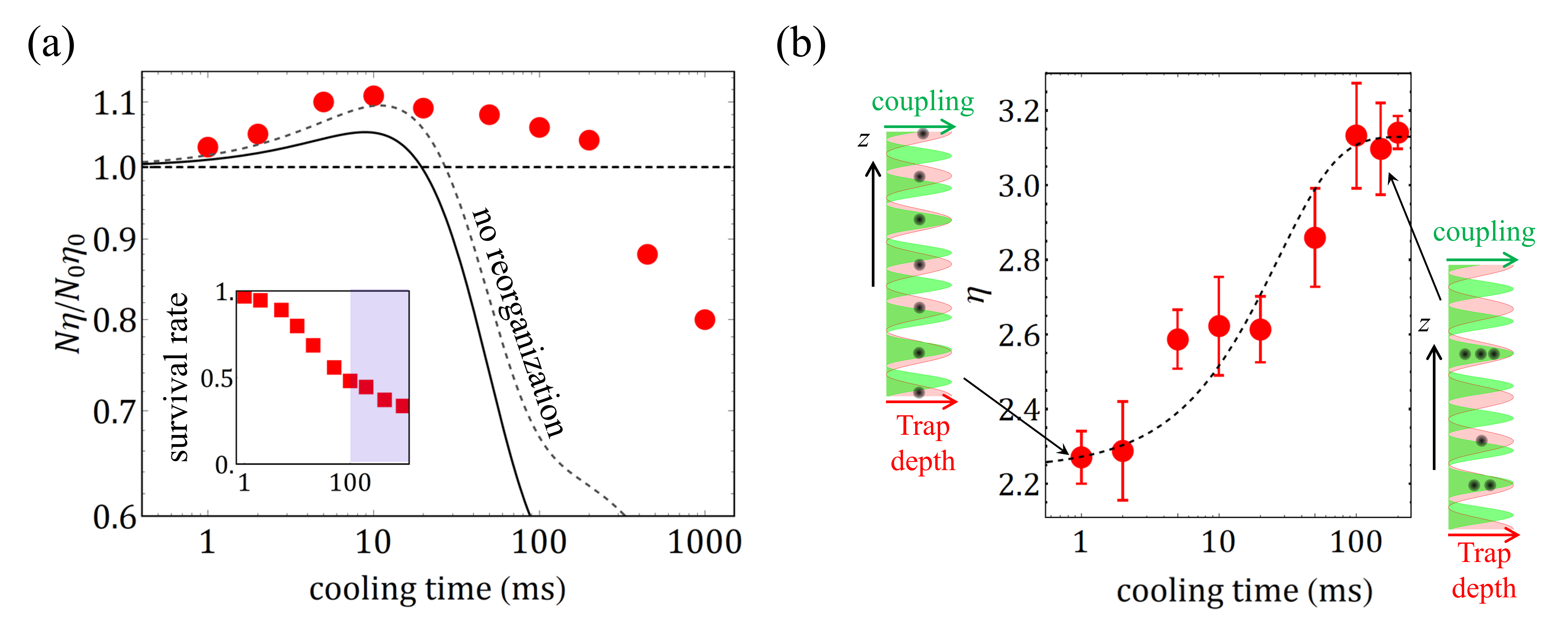}
\caption{Cavity coupling during the cooling process. (a) Collective cooperativity $N \eta$, normalized to the initial value, vs. cooling time. The two lines are obtained with a model based on the absence of atomic reorganization during the cooling process, i.e., weakly trapped atoms are lost. The solid line considers a transverse cooling-induced compression occurring along the weakly trapped direction, while the dashed line represents the case of an isotropic compression. The inset shows the inferred survival rate $N_\mathrm{tot}(t)/N_\mathrm{tot}(0)$ as a function of the cooling time $t$. Details and validation of the model are presented in the supplemental information \cite{SM}. (b) Effective cooperativity as a function of the cooling time. The dashed line is an exponential fit and serves as an eye guide. 
Insets: the schematic of atomic distribution among the incommensurate trapping and coupling lattices. Initially, atoms are evenly distributed in the trapping lattice, and after long-time cooling, the atoms are concentrated in lattice sites with high overlap with the coupling lattice. 
}\label{realFig5}
\end{figure*}

From Fig.~\ref{fig:Temperature}, we notice that for cooling times $t>30$ms the temperature changes much more slowly, and this is also true for the $\pi$-pulse fidelity after the corresponding cooling time. 
However, as we measure the effective atom number via the vacuum Rabi splitting of the cavity mode \cite{Braverman2019}, we find that the collective cooperativity $N \eta$ behaves quite differently from the temperature (see Fig.~\ref{realFig5}a). The collective cooperativity $N \eta$ increases at first, and later drops below its original value. Since the atom number $N$ can only decrease during the interaction with the cooling light, we conclude that the cavity coupling per atom (effective single-atom cooperativity $\eta$) must be increasing during the cooling. 

To extract the single-atom cooperativity, we then measure the quantum projection noise of a coherent spin state after cooling for different atom numbers (see Supplemental Material (SM) \cite{SM}). 
The spin quantum projection noise scales with the square root of the atom number $N$, while the total coupling strength $N \eta$ scales linearly with $N$. 
Therefore, comparing the spin noise variance with $N \eta$, the effective single-atom cooperativity $\eta$ can be deduced \cite{Schleier-Smith2010}. Fig.~\ref{realFig5}(b) shows that $\eta$ increases during the cooling until it saturates at 200~ms. Prior to the cooling process, the atoms are homogeneously distributed along the magic-wavelength lattice that has a different wavelength from the probing lattice near the $\ket{^1S_0} \rightarrow \ket{^3P_1}$ transition. The increase in single-atom cooperativity implies a redistribution of the atoms towards trapping sites with increased coupling to the probe light. (The radial cooling can also increase the single-atom cooperativity, but at most by 13\%, see SM \cite{SM}, while we observe a larger increase of 30\%.) If the increase in effective single-atom cooperativity beyond the transverse compression were due to the removal of weakly-coupled atoms as demonstrated in~\cite{wu2021sitedependent}, we would obtain the black solid line in Fig.~\ref{realFig5}(a), which disagrees with the data. The much higher remaining $N\eta$ in spite of atom loss requires a reorganization of atomic distribution along the cavity axis during cooling towards trapping sites with large coupling to the cavity probe light. 

We attribute the reorganization of the atoms along the cavity mode to the spatially dependent optical pumping and cooling that is performed with light in the same longitudinal and transverse cavity mode as the probe light. This means that atoms that are trapped in magic-wavelength lattice sites that are strongly coupled to the optical-pumping lattice experience strong cooling, while atoms that are trapped in sites that are weakly or not coupled to the optical pumping light are not cooled but experience photon recoil heating due to light scattering from the $\pi$-polarized Raman beam illuminating them from the side. Atoms that were originally loaded into such sites are then likely to be heated out, and can migrate to sites with good optical pumping and cooling, where they will be cooled deeply into the lattice. The latter sites are also strongly coupled to the probe light in the same mode, resulting in larger single-atom cooperativity. 
The observed time scale for reorganization of the atoms along the vertical cavity lattice of $\sim 50$~ms (see Fig. \ref{realFig5}) is much longer than the $\sim 10$~ms timescale for local cooling.

After 200~ms of cooling, the atomic cloud has rms sizes of $z_0=12\mu$m and $x_0= 3.2\mu$m, resulting in a peak occupation of $N_\mathrm{tube}=1.1$ atoms per tube. 
Even though the loaded atom number or atom survival during cooling were not optimized, the peak PSD of $1.3(3)\times 10^{-2}$ is already within a factor of 70 of quantum degeneracy (albeit currently at only one atom per tube). We believe that by using methods developed for Sr~\cite{Stellmer2013,Chen2022} and the alkalis~\cite{Hu2017,Solano2019,Vendeiro2022}, such as spectral shielding \cite{Stellmer2013} and spatial compression and recooling, it should be possible to reach quantum degeneracy by Raman sideband cooling in $^{171}$Yb. This would represent the first optical cooling to quantum degeneracy in a Fermi gas.
Even though the loaded atom number or atom survival during cooling were not optimized, the peak PSD of $1.3(3)\times 10^{-2}$ is already within a factor of 70 of quantum degeneracy (albeit currently at only one atom per tube). We believe that by using methods developed for Sr~\cite{Stellmer2013,Chen2022} and the alkalis~\cite{Hu2017,Solano2019,Vendeiro2022}, such as spectral shielding \cite{Stellmer2013} and spatial compression and recooling \cite{Hu2017}, it should be possible to reach quantum degeneracy by Raman sideband cooling in $^{171}$Yb. This would represent the first optical cooling to quantum degeneracy in a Fermi gas.

In conclusion, we have demonstrated Raman sideband cooling of nearly non-colliding atoms to near the motional ground state in two directions. 
This enables high-fidelity Rabi flopping on the optical-clock transition that is crucial for clock operation \cite{Blatt2009PRA,campbell2009probing,akatsuka2008optical} and precision beyond the standard quantum limit \cite{pedrozo2020entanglement,robinson2022direct}. 
In the future, a similar approach with improved optical access can likely be used to directly laser cool the fermionic gas to quantum degeneracy at small atom loss and in a cooling time substantially shorter than standard approaches with evaporative cooling.  

\bibliographystyle{apsrev4-2}
\bibliography{RamanCooling}
\newpage
\onecolumngrid
\section{Supplementary Information of ``Increased Atom-Cavity Coupling through Cooling-Induced Atomic Reorganization''}
\twocolumngrid
\renewcommand{\theequation}{S-\arabic{equation}}
\setcounter{figure}{0}
\renewcommand{\thefigure}{S-\arabic{figure}}

\subsection{Determine the Effective Single-Atom Cooperativity}

We characterize the single-atom cooperativity $\eta$ by measuring the spin projection noise via the cavity as a function of the collective cooperativity $N\eta$, where $N$ is the total number of atoms coupled to the cavity.
For a CSS prepared at the equator of the generalized Bloch sphere, the measured variance of the difference is 
\begin{equation}
\mathrm{var}\left(\sum_i\eta_i\sigma^i_z\right)= \frac{1}{4} N\langle\eta^2\rangle.
\end{equation}
We can thus obtain effective cooperativity by
\[\eta_\mathrm{eff}=\frac{\langle\eta^2\rangle}{\langle\eta\rangle}.\]

\subsection{Transversal Cooperativity Distribution}
We now consider the distribution of cooperativity due to the atomic position along the transversal (in the $xy$-plane) position. It's natural to consider a Gaussian atomic distribution with a waist of $\sigma_{x,y}$ for the $x,y$ directions respectively, where the anisotropicity comes from the different trapping frequencies. Then, the average cooperativity is 
\[\langle\eta\rangle=\int\mathrm{d}x\mathrm{d}yP_x(x)P_y(x)\eta(x,y),\]
where $P_x(x), P_y(y)$ stand for the Gaussian probability distribution of atoms and $\eta(x,y)=\eta_\mathrm{max}\exp\left(-2(x^2+y^2)/w_0^2\right)$ is the cooperativity for atom located at transversal position $(x,y)$, and $w_0=13.7\mu$m is the cavity mode waist that the atomic position. This means
\[\begin{split}
    \langle\eta\rangle&=\int\mathrm{d}x\mathrm{d}y\frac{1}{\sqrt{2\pi}\sigma_x}e^{-\frac{x^2}{2\sigma_x^2}}\frac{1}{\sqrt{2\pi}\sigma_y}e^{-\frac{y^2}{2\sigma_y^2}}\eta_\mathrm{max}e^{-2\frac{x^2+y^2}{w_0^2}}\\
    &=\eta_\mathrm{max}\frac{w_0^2}{\sqrt{(w_0^2+4\sigma_x^2)(w_0^2+4\sigma_y^2)}}. 
\end{split}\]

Similarly, we can obtain 
\[\begin{split}
    \langle\eta^2\rangle&=\int\mathrm{d}x\mathrm{d}y\frac{1}{\sqrt{2\pi}\sigma_x}e^{-\frac{x^2}{2\sigma_x^2}}\frac{1}{\sqrt{2\pi}\sigma_y}e^{-\frac{y^2}{2\sigma_y^2}}\eta_\mathrm{max}^2e^{-4\frac{x^2+y^2}{w_0^2}}\\
    &=\eta_\mathrm{max}^2\frac{w_0^2}{\sqrt{(w_0^2+8\sigma_x^2)(w_0^2+8\sigma_y^2)}}. 
\end{split}\]
This gives an effective cooperativity~\cite{Tanji-Suzuki2011,Hu2015}
\begin{align}
    \eta_\mathrm{eff}=\frac{\langle\eta^2\rangle}{\langle\eta\rangle}=\eta_\mathrm{max}\frac{\sqrt{(w_0^2+4\sigma_x^2)(w_0^2+4\sigma_y^2)}}{\sqrt{(w_0^2+8\sigma_x^2)(w_0^2+8\sigma_y^2)}}. 
\end{align}

\begin{figure}[!ht]
    \centering
    \includegraphics[width=.9\columnwidth]{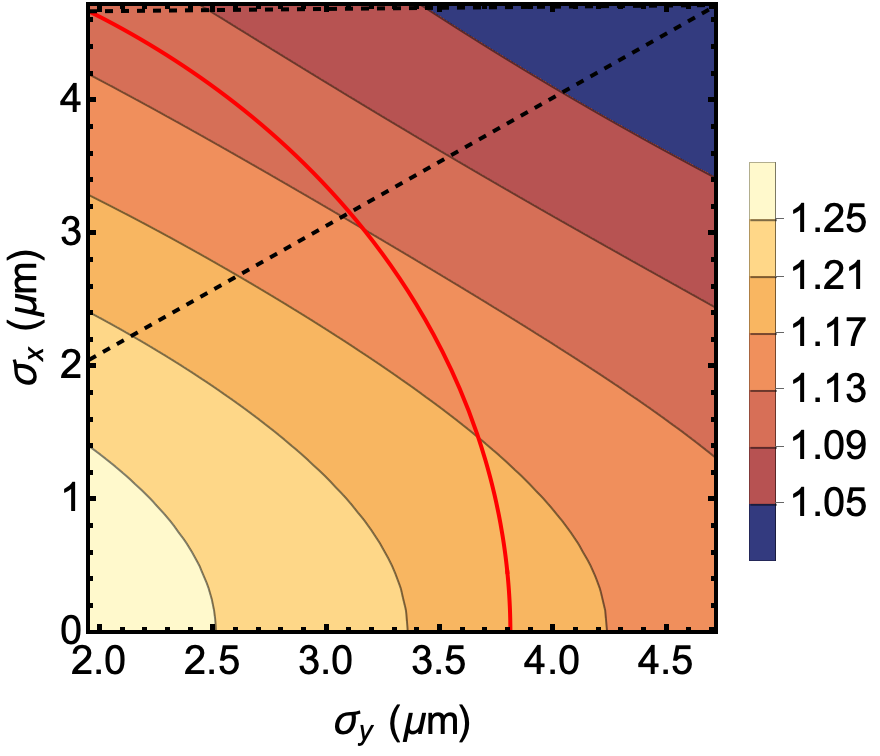}
    \caption{$\eta_\mathrm{eff}/\eta_\mathrm{eff,0}$ as a function of the final atom size $\sigma_{\{x,y\}}$.  }
    \label{fig:Supp1}
\end{figure}

We also know from the imaging that the atomic distribution before the cooling is isotropic with $\sigma_x=\sigma_y=4.7\mu$m. Then, we can obtain the after-cooling effective cooperativity $\eta_\mathrm{eff}$ as a function of the final atom size $\sigma_{\{x,y\}}$, as plotted in Fig.~\ref{fig:Supp1}.

From the imaging data, we know that the after-cooling gives a waist of projection along $\cos\theta\hat{y}-\sin\theta\hat{x}$ direction to be $3.1\mu$m, where $\theta=35^\circ$ is the imaging angle. In Fig.~\ref{fig:Supp1}, we plot the possible relation of $\sigma_x,\sigma_y$ to satisfy the final extension of $3.1\mu$m as the red curve. The relation is
\[\sigma_x=\sqrt{w^2-\sigma_y^2\cos^2\theta}/\sin\theta. \]
We also show an extreme isotropic case $\sigma_x=\sigma_y$ as a black dashed line, where the intersection of this line and the final allowed situation gives a 13\% increase in $\eta_\mathrm{eff}$ (the intersection of the two lines in Fig.~\ref{fig:Supp1}). 



\subsection{Phase-Space Density (PSD)}

\begin{figure}
    \centering
    \includegraphics[width=\columnwidth]{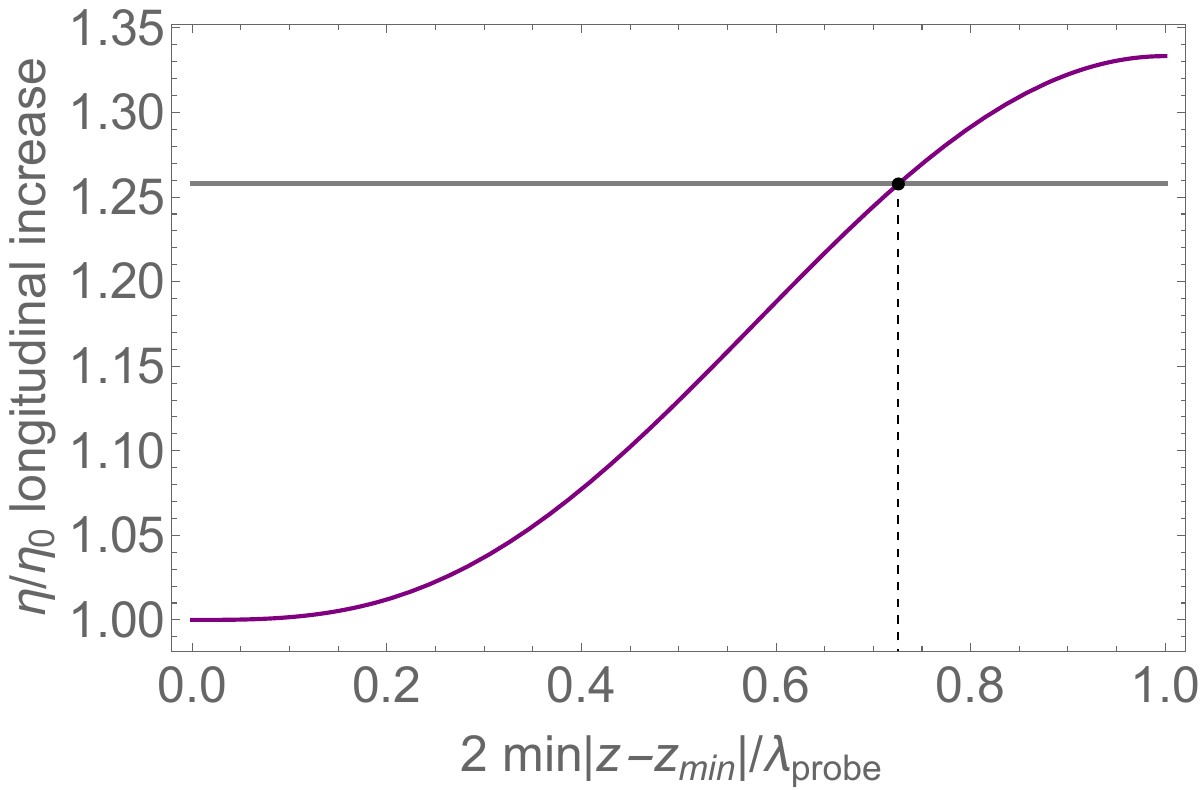}
    \caption{Increasing of $\eta$ as a function of cutoff. The gray solid line indicates the observed fractional $\eta$ increase, corresponding to a cutoff of all traps sitting at a distance $\leq 0.73 \lambda_\mathrm{probe}/2$ from a minimum of the cavity's longitudinal mode of the probe. }
    \label{fig:Supp2}
\end{figure}

Assume the spatial distribution of atoms is Gaussian on top of a 2D optical lattice. A useful number is the atoms per tube of the 2D lattice, which reaches its peak value at the 2D center of the cloud. We assume the Gaussian waist to be $\sigma_{x,y,z}$ along the three directions where $y$ is the elongated tube direction. Then, the normalized peak 2D distribution is $p_\mathrm{2D}(\sigma_x,\sigma_z)=1/(2\pi\sigma_x\sigma_z)$, and the peak number of atoms per tube is
\[N_\mathrm{tube}=N\cdot p_\mathrm{2D}(\sigma_x,\sigma_z)\left(\frac{\lambda_\mathrm{trap}}{2}\right)^2, \]
where $N$ is the atom number, and $\lambda_\mathrm{trap}=759$nm is the magic trap's wavelength. We can thus calculate the atoms per tube before and after the cooling, given the atom numbers are $N_\mathrm{hot}=1300$ and $N_\mathrm{cold}=641$. However, due to the reorganization, some tubes are no longer occupied and thus, a factor $q$ should be inserted to correct this effect for the cold case. To match the increase of the cooperativity, we find that 
$\approx 73\%$ of vertical traps should be empty (see Fig.~\ref{fig:Supp2}), making
\[q = \frac{1}{1-0.73}.\]
Thus, the atoms per tube at peak is 0.4 for hot atoms and 1.13 for cold atoms. 

In order to derive the PSD, we need to further calculate the thermal occupation of the vibrational levels as
\[\mathrm{PSD}=N_\mathrm{tube}n_xn_yn_z,\]
where $n_{x,y,z}=1-\exp(-(\hbar\omega_{x,y,z})/k_BT)$ with $\omega_{x,y,z}=2\pi\times\{62,1,140\}$kHz, respectively. Specifically, the trapping frequency along the cavity direction $\omega_z$ is already sideband resolved for the $\ket{g}\leftrightarrow\ket{e}$ transition, making the distribution $n_z$ the same before and after the cooling. Altogether, we have the PSD to be 0.00008(2) before the cooling and 0.013(3) after the cooling. 
\end{document}